\documentclass[a4,11pt,reqno]{amsart}%
\usepackage[english]{babel}
\usepackage[latin1]{inputenc}
\usepackage{graphics}
\usepackage{ulem}
\usepackage{hhline}
\usepackage{dsfont}
\usepackage{mathrsfs}
\usepackage{fancyhdr}
\usepackage{amsmath,amssymb}
\usepackage{rotating}
\usepackage[latin1]{inputenc}
\usepackage[T1]{fontenc}
\usepackage{fancybox}
\usepackage{color}
\usepackage{colortbl}
\usepackage{setspace}
\usepackage{enumerate}
\usepackage[nice]{nicefrac}
\usepackage{amsthm}
\usepackage{multicol}
\usepackage{pifont}
\usepackage[french]{minitoc}
\usepackage{latexsym,amsfonts}
\usepackage{amsthm}
\usepackage{varioref}
\usepackage{textcomp}
\usepackage{lmodern}
\usepackage{mathpazo}
\usepackage{euscript}
\usepackage[pdftex]{hyperref}
\numberwithin{equation}{section}
\makeatletter\@addtoreset{equation}{section}

\DeclareMathSymbol{\subsetneqq}{\mathbin}{AMSb}{36}
\newtheorem {theorem}{Theorem}[section]

\newtheorem {proposition}[theorem]{Proposition}

\newcommand{\C}{\mathbb C}
\newcommand{\R}{\mathbb R}

\begin{document}

\title[Magnetic Berezin transforms as functions of the
Laplacian]{A Formula representing magnetic Berezin transforms as functions of the
Laplacian on $\C^n$}

\author[N. Askour, A. Intissar,  Z. Mouayn]{{\bf  Nour Eddine Askour, Ahmed Intissar and Zouha\"{i}r Mouayn}}
\address{ {\sf Nour Eddine Askour $\&$ Zouha\"{i}r Mouayn} \newline
        {Department of Mathematics, Faculty of Sciences and Technics (M'Ghila), PO. Box 523,
          \newline
          Sultan My Slimane University, CP. 23 000, B\'{e}ni Mellal, Morocco}}
          \email{askour2a@fstbm.ac.ma} \email{mouayn@fstbm.ac.ma}
\address{{ \sf Ahmed Intissar} \newline
        {Department of Mathematics, Faculty of Sciences,
        \newline
        P.O. Box 1014, Mohammed V University, Agdal, 10000 Rabat, Morocco}}
        \email{intissar@fsr.as.ma}


\date{}
\maketitle


\begin{abstract}
we give a formula that express magnetic Berezin transforms associated with
generalized Bargmann-Fock spaces as a functions of the Euclidean Laplacian
on $\C^n.$

%
\end{abstract}

\section{Introduction}

The Berezin transform was introduced by Berezin \cite{1} for certain classical
symetric domains in $\C^n$. This transform links the Berezin symbols
and the symbols for Toeplitz operators. It is present in the study of the
correspondence principle. The formula representing the Berezin transform as
function of the Laplace-Beltrami operator plays a key role in the Berezin
quantization \cite{1}.

This transform can be defined as follows. Consider a domain $D\subseteq \C^n$ and a Borel measure on $D$. Let $\mathcal{H}$ be a closed subspace
of $L^{2}(D,d\mu) $ consisting of continuous functions and we
assume that $\mathcal{H}$ has a reproducing kernel $K( .,.) $.
Then, The Berezin symbol $\sigma (A) $ of a bounded linear
operator $A$ on $\mathcal{H}$ is the function on $D$ given by $\sigma (A) (z) =\left\langle Ae_{z},e_{z}\right\rangle $, 
where $e_{z}(.) =K(z,z) ^{-\frac{1}{2}}K(.,z) $ $\in \mathcal{H}$. For each $\varphi \in L^{\infty }(D) $, the Toeplitz operator $T_{\varphi }$ with symbol $\varphi $ is
the operator on  $\mathcal{H}$ given by $T_{\varphi }\left[
f\right] =P( \varphi f) $, $f\in \mathcal{H}$ where $P$ is the
orthogonal projection from $L^{2}(D,d\mu) $ into $\mathcal{H}$.
The Berezin transform associated to $\mathcal{H}$ is, by definition, the
positive self-adjoint operator $\sigma (T) $, which turns out to
be a bounded operator on $L^{2}(D,d\mu) $, where $d\mu_{K}=K(z,z) d\mu (z) .$

Now, based on the consideration that the Berezin transform can be defined
provided that there is a given closed subspace $L^{2}$ which possesses a
reproducing kernel, we are here concerned with the eigenspaces
\begin{equation}
A_{m}^{2}(\C^n) =\left\{ \psi \in L^{2}(\C^n,e^{-|z|^{2}}d\mu), \quad \widetilde{\Delta }\psi
=\epsilon _{m}\psi \right\}   \label{1.1}
\end{equation}
of the second order differential operator
\begin{equation}
\widetilde{\Delta }=-\sum\limits_{j=1}^n\frac{\partial ^{2}}{\partial
z_{j}\partial \overline{z}_{j}}+\sum\limits_{j=1}^n\overline{z}_{j}\frac{%
\partial }{\partial \overline{z}_{j}},  \label{1.2}
\end{equation}
corresponding to eigenvalues $\epsilon _{m}=m$, $m=0,1,2,\ldots $. Here $d\mu $
is the Lebesgue measure on $\C^n$. The operator $\widetilde{\Delta }$
constitutes (in suitable units and up to an additive constant), in $%
L^{2}( \C^n,e^{-|z|^{2}}d\mu) $, a
realization of the Schr\"{o}dinger operator with uniform magnetic field in $%
\C^n$. Theses eigenspaces are reproducing kernel Hilbert spaces with
reproducing kernels given by (\cite{2}):
\begin{equation}
K_{m}(z,w) :=\pi ^{-n}e^{\left\langle z,w\right\rangle
}L_{m}^{(n-1) }\left( |z-w|^{2}\right) ,w,z\in \C^n,  \label{1.3}
\end{equation}
where $L_{k}^{(\alpha) }(x) $ is the Laguerre polynomial \cite[p. 239]{3}.

Actually, by \cite{2} it is known that the eigenspace $A_{0}^{2}(\C^n) $    corresponding to  $m=0$
coincides with Bargmann-Fock $\mathcal{F}(\C^n) $ space
of holomorphic functions that are $e^{-|z|^{2}}d\mu $- square
integrable, while for $m\neq 0$, the spaces $A_{m}^{2}(\C^n) $ which can be viewed as kernel spaces of the hypoelliptic
differential operator $\left( \widetilde{\Delta }-m\right) $, consist of non
holomorphic functions, These spaces are called generalized Bargmann-Fock spaces.

Note also, for $m=0$, the Berezin transform, denoted $B_{0}$, associated
with the Bargmann-Fock space $\mathcal{F}(\C^n) $ turns
out to be given by a convolution product over the group $\C^n=\R^{2n}$ as 
$$B_{0}[\phi] (z) :=\left( \pi^{-n}e^{-|w|^{2}}*\phi \right) (z) , \qquad \phi \in L^{2}( \C^n,d\mu ) .$$ Furthermore, it can be expressed as
a function of the Euclidean Laplacian on $\C^n$ as $B_{0}=e^{\frac{1}{4}\Delta _{\C^n}}$ \cite{4}.

In this paper, we extend to each eigenspace $A_{m}^{2}(\C^n) $ the notion of Berezin transform by considering the transform
defined via the following convolution product over $\C^n$ as
\begin{equation}
B_{m}[\phi] (z) :=\frac{\pi ^{-n}m!}{(n) _{m}}\left( e^{-|w|^{2}}\left( L_{m}^{(n-1) }( |w|^{2}) \right)^{2}*\phi \right)
(z) ,\phi \in L^{2}(\C^n,d\mu) ,
\label{1.4}
\end{equation}
and we prove that this transform can also be expressed as a
function of the Laplacian $\Delta _{\C^n}$ as:
\begin{equation}
B_{m}:=\frac{1}{(n) _{m}}e^{\frac{1}{4}\Delta _{\C%
^n}}\sum\limits_{k=0}^{m}\frac{(n-1) _{k}\left( m-k\right) !}{%
k!}\left( \frac{\Delta _{\C^n}}{4}\right) ^{k}L_{m-k}^{\left(
k\right) }\left( \frac{\Delta _{\C^n}}{4}\right) L_{m-k}^{\left(
n-1+k\right) }\left( \frac{\Delta _{\C^n}}{4}\right) ,  \label{1.5}
\end{equation}
where $(\alpha) _{j}=\alpha (\alpha +1) \cdots ( \alpha +j-1) $ denotes the Pochhammer symbol.

This paper is organized as follows. In Section 2 we recall same needed facts
on the generalized Bargmann-Fock spaces. In Section 3, we apply the
formalism of the Berezin transform so as to extend this notion to each
generalized Bargmann-Fock spaces. In Section 4, we give a formula that
represents the extended Berezin transform as a function of the Laplacian in
the Euclidean complex $n$-space .

\section{The Schr\"{o}dinger operator with magnetic field on $\C^n$.}

The motion of charged particle in a constant uniform magnetic field in $\R^{2n}$ is described 
(in suitable units and up to additive constant) by the
Schr\"{o}dinger operator:
\begin{equation}
H_{B}:=-\frac{1}{4}\sum\limits_{j=1}^n\left( \partial
_{x_{j}}+By_{j}\right) ^{2}+\left( \partial _{y_{j}}-iBx_{j}\right) ^{2}-%
\frac{n}{2}  \label{2.1}
\end{equation}
acting on $L^{2}\left( \R^{2n},d\mu \right) $, $B>0$ is a constant
proportional to the magnetic field strength. We identify the Euclidean space
$\R^{2n}$ with $\C^n$ in the usual way. The operator $H_{B}$ in
equation (2.1) can be represented by the operator
\begin{equation}
\widetilde{H}_{B}=e^{\frac{1}{2}B|z|^{2}}H_{B}e^{-\frac{1}{2}B|z|^{2}}.  \label{2.2}
\end{equation}
According to equation \eqref{2.2},  an arbitrary state $\phi $ of $L^{2}( \R^{2n},d\mu ) $  is represented by the
function $Q[\phi] $ of $L^{2}( \C^n,e^{-|z| ^{2}}d\mu) $defined by
\begin{equation}
Q[\phi] (z) :=e^{\frac{1}{2}|z|^{2}}\phi (z) ,z\in \C^n.  \label{2.3}
\end{equation}
The unitary map $Q$ in \eqref{2.3} is called ground state transformation. For $B=1$, 
the explicit expression for the operator in equation \eqref{2.2} turns out to be
given by the operator $\widetilde{\Delta }$ introduced in equation \eqref{1.2}.
The Latter is considered with $C_{0}^{\infty }(\C^n) $ as
its regular domain in the Hilbert space $L^{2}(\C^n,e^{-|z| ^{2}}d\mu) $ of $e^{-|z|^{2}}d\mu $-square
integrable functions $\varphi :\C^n\rightarrow \C$.

We let $P_{m}:L^{2}( \C^n,e^{-|z|^{2}}d\mu)
\rightarrow A_{m}^{2}(\C^n) $ denote the orthogonal
projection operator onto the eigenspace $A_{m}^{2}(\C^n) $
as defined in \eqref{1.1}. A basis elements of this space can be written
explicitly in terms of the Laguerre polynomials $L_{k}^{(\alpha) }(x) $ and the polynomials $h_{p,q}^{j}( z,\overline{z}) $ whose are the restriction to the unit sphere $S^{2n-1}$
of harmonic homogeneous polynomials of bidegree $( p,q) $
\cite[p. 253]{5}. Precisely, the following set of the functions
\begin{equation}
\Psi _{j,p,q}^{m}(z) =\left( \frac{2( m-q) !}{\Gamma
( n+m+p) }\right) ^{\frac{1}{2}}L_{m-q}^{(n+p+q-1)}\left( |z|^{2}\right) h_{p,q}^{j}( z,\overline{z})
\label{2.4}
\end{equation}
constitutes an orthonormal basis of $A_{m}^{2}(\C^n) $,
for varying $p=0,1,2,\ldots ;q=0,1,\ldots ,m$ and $j=1,\ldots ,d_{p,q}^n$
with $d_{p,q}^n=\dim \mathcal{H}_{p,q}\left( S^{2n-1}\right) $, where $%
\mathcal{H}_{p,q}\left( S^{2n-1}\right) $ is finite dimensional vector space
spanned by the above harmonic polynomials $h_{p,q}^{j}$. The above basis can
be used to obtain the reproducing kernel of the Hilbert space $%
A_{m}^{2}(\C^n) $ by following general theory \cite{6}.
Actually, as mentioned in section 1, this kernel is of the form
\begin{equation}
K_{m}(z,w) =\pi ^{-n}e^{\left\langle z,w\right\rangle
}L_{m}^{(n-1) }\left( |z-w|^{2}\right) ,z,w\in \Bbb{%
C}^n  \label{2.5}
\end{equation}
For more informations on the spectral properties of  the operator
 $\widetilde{\Delta }$ and its eigenspaces $A_{m}^{2}\left( \Bbb{C%
}^n\right) $ we refer to \cite{7}.

\textbf{Remarque 2.1} Note that, for $m=0$, the kernel $K_{0}\left(
z,w\right) =\pi ^{-n}e^{\left\langle z,w\right\rangle }$ coincides with the
reproducing kernel of Bargmann-Fock $\mathcal{F}(\C^n) $.

\section{Magnetic Berezin Transforms}

According to the formalism described in Section 1, we take as domain $D=\Bbb{%
C}^n$ the whole complex space. For a bounded operator $A$ on 
$\mathcal{H}:= A_{m}^{2}(\C^n) \subset L^{2}(\C^n,e^{-|z|^{2}}d\mu) $, the Berezin symbol of $A$ is%
  defined by
\begin{equation}
\sigma _{m}(A) :=\left\langle Ae_{z,m},e_{z,m}\right\rangle _{%
\mathcal{H}}  \label{3.1}
\end{equation}
where $e_{z,m}(.) :=( K_{m}(z,z)) ^{-\frac{1}{2}}K_{m}( z,.) $ denotes the normalized reproducing
kernel according to \eqref{2.5} of $A_{m}^{2}(\C^n) $ with
evaluation at $z\in \C^n$, precisely,
\begin{equation}
e_{z,m}(w) =\pi ^{\frac{n}{2}}\left( \frac{m!}{(n)_{m}}\right) ^{-\frac{1}{2}}e^{-\frac{|z|^{2}}{2}}K_{m}(z,w) ,\quad w\in \C^n.  \label{3.2}
\end{equation}
For a bounded function on $\C^n$, the Toeplitz operator $T_{\varphi }$
is the operator $T_{\varphi }(h) =P_{m}(\varphi h), h\in L^{2}(\C^n,e^{-|z|^{2}}d\mu) $. the
Berezin transform of the function $\varphi $ is defined to be the Berezin
symbol $\sigma _{m}\left( T_{\varphi }\right) $. That is,
\begin{equation}
B_{m}[\varphi] (z) :=\sigma _{m}\left( T_{\varphi}\right) (z) =\left\langle T_{\varphi }\left( e_{z,m}\right)
,e_{z,m}\right\rangle _{_{\mathcal{H}}},z\in \C^n  \label{3.3}
\end{equation}
Explicitly, this transform reads
\begin{equation}
B_{m}[\varphi] (z) =\frac{m!}{(n)
_{m}\pi ^n}\int_{\C^n}e^{-|z-w|^{2}}\left(L_{m}^{(n-1) }\left( |z-w|^{2}\right) \right)
^{2}\varphi (w) d\mu (w) ,   \label{3.4}
\end{equation}
where $\varphi \in L^{\infty }(\C^n) $.

As mentioned in the introduction, is easy to see from \eqref{3.4} that the
transform $B_{m}$, can written as a convolution operator as
\begin{equation}
B_{m}[\varphi] =b_{m}*\varphi ,\quad \varphi \in L^{2}(\C^n,d\mu)  \label{3.5}
\end{equation}
where
\begin{equation}
b_{m}(z) =\frac{m!}{(n) _{m}\pi ^n}e^{-|z| ^{2}}\left( L_{m}^{(n-1) }( |z|^{2}) \right) ^{2},z\in \C^n.  \label{3.6}
\end{equation}
Is not difficult to see that the function $b_{m}$ belongs to $L^{1}\left(
\C^n\right) $ by making use of the orthogonality relation of Laguerre
polynomials \cite[p. 241]{3}. Indeed, $\left\| b_{m}\right\| _{L^{1}(\C^n) }=\pi ^{-n}$. Then, applying Hausdorff-Young inequality to $b_{m}*\varphi $  unable us to write that
\begin{equation}
\left\| B_{m}[\varphi] \right\| _{L^{2}(\C^n) }\leq \pi ^{-n}
\left\| \varphi \right\|_{L^{2}(\C^n) }  \label{3.7}
\end{equation}
which means that $B_{m}:L^{2}(\C^n) \rightarrow L^{2}(\C^n) $ is a bounded operator and such as we will
be dealing with.

\smallskip

\textbf{Remark 3.1}. Note that, a decomposition of the action of $B_{m}$ on
the product of radial functions with spherical harmonics have been discussed
in \cite{8}.

\section{Berezin transform and the Euclidean Laplacian}

In this section, we shall express the transform $B_{m}$ as function of the
Euclidean Laplacian of $\C^n$. For this, we first state the following:

\begin{proposition}
Let $m\in \Bbb{Z}_{+}$, then there exists a function $f_{m}$ such that $%
B_{m}=f_{m}\left( \Delta _{\C^n}\right) ,$ with $f_{m}\left( \lambda
\right) =e^{-\frac{\lambda }{4}}P_{m}(\lambda) $, $P_{m}(.) $ being a polynomial function.
\end{proposition}

\proof
Since the transform $B_{m}$ can be written in view of \eqref{3.5} as a convolution
over $\C^n$ as $B_{m}[\varphi] =b_{m}*\varphi $, $%
\varphi \in L^{2}(\C^n) ,$ the function $b_{m}$ being
defined in \eqref{3.6}, then by the general theory (\cite[p. 200]{9}) we can write that:
\begin{equation}
B_{m}=\widehat{b}_{m}\left( \frac{1}{i}\nabla \right)  \label{4-1}
\end{equation}
where
\begin{equation}
\widehat{b_{m}}(\xi) =\int_{\C^n}e^{-i(\xi
\mid w) }b_{m}(w) d\mu (w)  \label{4.2}
\end{equation}
is the Fourier transform of the function $b_{m}$. Here, $( \xi \mid w) $ denotes the real scalar product in $\C^n=\R^{2n}$ and
$\nabla $ is the gradient operator on $\R^{2n}$. Using the expression
of the Laguerre polynomial (\cite[p. 240]{3}):
\begin{equation}
L_{m}^{(n-1) }(x) =\sum\limits_{k=0}^{m}(-1) ^{k}\binom{m+n-1}{m-k}\frac{x^{k}}{k!},  \label{4-3}
\end{equation}
then, from \eqref{4.3} we can write
\begin{equation}
L_{m}^{(n-1) }\left( |w|^{2}\right)
=\sum\limits_{k=0}^{m}c_{k}|w|^{2k},\quad c_{k}=\frac{\left(
-1\right) ^{k}}{k!}\binom{m+n-1}{m-k}.  \label{4.4}
\end{equation}

Inserting in \eqref{4.2} the expression of the Laguerre polynomial given in \eqref{4.4}, we obtain
\begin{align}
\widehat{b}_{m}\left( \xi \right) &=\frac{m!}{\pi ^n(n) _{m}}%
\sum\limits_{k=0}^{m}\sum\limits_{j=0}^{m}c_{j}c_{k}\int_{\C%
^n}e^{-|w|^{2}}e^{-i\left( \xi \mid w\right) }\left| w\right|
^{2\left( j+k\right) }d\mu (w)
\nonumber \\
&=\frac{m!}{\pi ^n(n) _{m}}\sum\limits_{k=0}^{m}\sum%
\limits_{j=0}^{m}c_{j}c_{k}\left( \Delta _{\xi }\right) ^{j+k}\left(
\int_{\C^n}e^{-|w|^{2}}e^{-i\left( \xi \mid w\right) }d\mu (w) \right)  \label{4.5}
\end{align}
where $\Delta _{\xi }=4\sum\limits_{j=1}^n\frac{\partial ^{2}}{\partial
\xi _{j}\partial \overline{\xi }_{j}}$ is the Laplacian in terms of the
variable $\xi $.

The last integral is recognized as the Gaussian integral (\cite[p. 153]{10}):
\begin{equation}
\int_{\C^n}e^{-|w|^{2}}e^{-i\left( \xi \mid w\right) }d\mu (w) =\pi ^ne^{-\frac{|\xi|^{2}}{4}%
}  \label{4.6}
\end{equation}
Thus, making use of \eqref{4.6}, equation \eqref{4.5} reads
\begin{equation}
\widehat{b}_{m}\left( \xi \right) =\frac{m!}{(n) _{m}}%
\sum\limits_{k=0}^{m}\sum\limits_{j=0}^{m}c_{j}c_{k}\Delta _{\xi
}^{j+k}\left( e^{-\frac{|\xi|^{2}}{4}}\right) .  \label{4.7}
\end{equation}
Now, the last term in \eqref{4.7} can be expressed as
\begin{equation}
\Delta _{\xi }^{j+k}\left( e^{-\frac{|\xi|^{2}}{4}}\right)
=P_{j,k}\left( |\xi|^{2}\right) e^{-\frac{\left| \xi \right|
^{2}}{4}}.  \label{4.8}
\end{equation}
where $P_{j,k}(.) $ is polynomial function. Therefore, equation
\eqref{4.7} becomes
\begin{equation}
\widehat{b}_{m}(\xi) =\frac{m!}{(n) _{m}}%
\sum\limits_{k=0}^{m}\sum\limits_{j=0}^{m}c_{j}c_{k}P_{j,k}\left(|\xi|^{2}\right) e^{-\frac{|\xi|^{2}}{4}}=e^{-\frac{|\xi|^{2}}{4}}P_{m}(|\xi| ^{2})   \label{4.9}
\end{equation}
Finally, replacing $\xi $ by $\frac{1}{i}\nabla $, we arrive at announced
statement of Proposition 1. This ends the proof.

\begin{theorem}
Let $m\in \Bbb{Z}_{+}$. Then, the Berezin transform $B_{m}$ can be expressed
 in terms of the Laplacian $\Delta _{\C^n}$ as
\begin{equation}
B_{m}=\frac{1}{(n) _{m}}\exp \left( \frac{\Delta _{\C^n}}{4}\right) \sum\limits_{k=0}^{m}\frac{(n-1)_{k}(m-k)
!}{k!}\left( \frac{\Delta _{\C^n}}{4}\right) ^{k}L_{m-k}^{(k) }\left( \frac{-\Delta _{\C^n}}{4}\right) L_{m-k}^{(n-1+k) }\left( \frac{-\Delta _{\C^n}}{4}\right)   \label{4.10}
\end{equation}
\end{theorem}
Using of Proposition 1 together with general theory of the function of
self-adjoint operator (\cite[p. 338]{11}), we can link the Berezin transform $B_{m}$ with the spectral family 
$\left\{ E_{\lambda },\lambda >0\right\} $ associated to $-\Delta _{\C^n}$ as
\begin{equation}
\left\langle B_{m}\varphi ,\psi \right\rangle _{L^{2}(\C^n) }=\left\langle f_{m}\left( -\Delta _{\C^n}\right) \varphi
,\psi \right\rangle _{L^{2}(\C^n)
}=\int_{0}^{+\infty }f_{m}(\lambda) ,d\left\langle
E_{\lambda }[\varphi] ,\psi \right\rangle _{L^{2}(\C^n) }  \label{4.11}
\end{equation}
$\varphi ,\psi \in L^{2}(\C^n) $ and $E_{\lambda }$ is
the well known spectral projector given by (\cite[9p. 202]{9}):
\begin{equation}
E_{\lambda }\varphi (z) =\frac{\lambda ^{\frac{n}{2}}}{\left(
2\pi \right) ^n}\int_{\C^n}\left| z-w\right|
^{-n}J_{n}\left( \lambda ^{\frac{1}{2}}|z-w|\right) \varphi
(w) d\mu (w) ,  \label{4.12}
\end{equation}
$J_{\nu }(.) $ being the Bessel function of order $\nu $
(\cite[p. 65]{3}).

Using an integration by part in the Stieltjes' sense, we obtain from \eqref{4.11}
\begin{align}
\left\langle B_{m}\varphi ,\psi \right\rangle _{L^{2}(\C^n) }&=-\int_{0}^{+\infty }\frac{df_{m}(\lambda) }{%
d\lambda }\left\langle E_{\lambda }\varphi ,\psi \right\rangle _{L^{2}(\C^n) }d\lambda  \label{4.13}
\\
&=\left\langle \int_{0}^{+\infty }-\frac{df_{m}(\lambda)
}{d\lambda }E_{\lambda }[\varphi] d\lambda ,\psi \right\rangle
_{L^{2}(\C^n) },\forall \psi \in L^{2}(\C^n,d\mu)  \label{4.14}
\end{align}
This relation \eqref{4.14} implies that
\begin{equation}
B_{m}[\varphi] =\int_{0}^{+\infty }-\frac{df_{m}(\lambda) }{d\lambda }E_{\lambda }[\varphi] d\lambda
\text{,\quad }\varphi \in L^{2}(\C^n,d\mu)  \label{4.15}
\end{equation}
which can also be written, in view of \eqref{4.12}, as:
\begin{equation}
B_{m}[\varphi] (z) =\frac{1}{\left( 2\pi \right)
^n}\int_{0}^{\infty }\frac{-df_{m}(\lambda) }{%
d\lambda }\lambda ^{\frac{n}{2}}\int_{\C^n}\frac{J_{n}\left(
\lambda ^{\frac{1}{2}}|z-w|\right) }{|z-w|^n}%
\varphi (w) d\mu (w) d\lambda  \label{4.16}
\end{equation}
\begin{equation}
=\int_{\C^n}\frac{-1}{\left( 2\pi \right) ^n}\left(
\int_{0}^{+\infty }\lambda ^{\frac{n}{2}}\frac{df_{m}\left( \lambda
\right) }{d\lambda }\frac{J_{n}\left( \lambda ^{\frac{1}{2}}\left|
z-w\right| \right) }{|z-w|^n}d\lambda \right) \varphi \left(
w\right) d\mu (w) .  \label{4.17}
\end{equation}
On the other hand, recalling the expression of $B_{m}$ given in \eqref{3.4} as:
\begin{equation}
B_{m}[\varphi] (z) =\frac{m!}{(n)
_{m}\pi ^n}\int_{\C^n}e^{-|z-w|^{2}}\left(
L_{m}^{(n-1) }( |z-w|^{2}) \right)
^{2}\varphi (w) d\mu (w) .  \label{4.18}
\end{equation}
We are lead to consider the following equality:
\begin{equation}
\frac{-1}{(2\pi) ^n}\int_{0}^{+\infty }\left( \lambda
^{\frac{n}{2}}\frac{df_{m}(\lambda) }{d\lambda }\frac{%
J_{n}\left( \lambda ^{\frac{1}{2}}|z-w|\right) }{|z-w| ^n}\right) d\lambda =\frac{m!}{(n) _{m}\pi ^n}%
e^{-|z-w|^{2}}\left( L_{m}^{(n-1) }( |z-w|^{2}) \right) ^{2}  \label{4.19}
\end{equation}
The equation \eqref{4.19} can be also written as
\begin{equation}
\int_{0}^{+\infty }\lambda ^{\frac{n}{2}}g_{m}(\lambda)
J_{n}\left( \lambda ^{\frac{1}{2}}x\right) d\lambda
=-C_{n,m}x^ne^{-x^{2}}\left( L_{m}^{(n-1) }(x^2)
\right) ^{2}  \label{4.20}
\end{equation}
where $g_{m}(\lambda) =$ $\frac{df_{m}(t) }{dt}\mid
_{t=\lambda }$, $x=|z-w|$ and $C_{n,m}=\frac{2^nm!}{(n) _{m}}$.

Changing the variable of integration by writing $\lambda =s^{2}$, we get
from (4.20)
\begin{equation}
\int_{0}^{+\infty }s^ng_{m}(s^{2}) sJ_{n}(sx) ds=\frac{1}{2}C_{n,m}x^ne^{-x^{2}}\left( L_{m}^{(n-1) }(x^2) \right) ^{2}.  \label{4.21}
\end{equation}
The left hand side of equation \eqref{4.21} can be presented as Hankel transform as
\begin{equation}
\mathcal{H}_{n}\left[ s\mapsto s^ng_{m}( s^{2}) \right] (x) =\frac{1}{2}C_{n,m}x^ne^{-x^{2}}\left( L_{m}^{(n-1)
}(x^2) \right) ^{2}  \label{4.22}
\end{equation}
where $\mathcal{H}_{n}$ is defined by (see \cite[p. 67]{12}):
\begin{equation}
\mathcal{H}_{n}[u(s)] (x)=\int_{0}^{+\infty }u(s) sJ_{n}(sx) ds.
\label{4.23}
\end{equation}
and satisfies the involution property
\begin{equation}
\int_{0}^{+\infty }\left( \int_{0}^{+\infty }u(x)J_{\nu }\left( tx\right) xdx\right) J_{\nu }( ts) tdt=u(s)  \label{4.24}
\end{equation}
which holds for every continuous function $u$ on $] 0,+\infty[ $
with$\int_{0}^{+\infty }x^{\frac{1}{2}}|u(x)| dx<\infty )$.

Making use of the involution property \eqref{4.24} for $\nu =n$, $u(s)
=s^ng_{m}\left( s^{2}\right) $, the equation \eqref{4.22}  becomes
\begin{equation}
s^ng_{m}\left( s^{2}\right) =-\frac{1}{2}C_{n,m}\int_{0}^{+\infty
}x^{n+1}e^{-x^{2}}\left( L_{m}^{(n-1) }(x^2)\right) ^{2}J_{n}(xs) dx.  \label{4.25}
\end{equation}
This, implies that the function $g_{m}(\lambda) $ is of the
form
\begin{equation}
g_{m}(\lambda) =-\frac{1}{2}C_{n,m}\int_{0}^{+\infty}x^{n+1}e^{-x^{2}}\left( L_{m}^{(n-1) }(x^2)\right) ^{2}
\lambda ^{-\frac{n}{2}}J_{n}\left( x\lambda ^{\frac{1}{2}}\right) dx=\frac{df_{m}(\lambda) }{d\lambda }.
\label{4.26}
\end{equation}
Let $t>0$, then by an integration over the interval $] t,+\infty[ $, we have
\begin{equation}
\lim\limits_{\lambda \rightarrow +\infty }f_{m}(\lambda)
-f_{m}(t) =-\frac{1}{2}C_{n,m}\int_{0}^{+\infty
}x^{n+1}e^{-x^{2}}\left( L_{m}^{(n-1) }(x) \right)
^{2}\int_{t}^{+\infty }\lambda ^{-\frac{n}{2}}J_{n}\left( x\lambda ^{%
\frac{1}{2}}\right) d\lambda dx.  \label{4.27}
\end{equation}
Taking into account the form of the function $f_{m}$ in the proposition 1, we
have
\begin{equation}
\lim\limits_{\lambda \rightarrow +\infty }f_{m}(\lambda)
=\lim\limits_{\lambda \rightarrow +\infty }e^{-\frac{\lambda }{4}}P_{m}(\lambda) =0.  \label{4.28}
\end{equation}
Then, we obtain:
\begin{equation}
f_{m}(t) =\frac{1}{2}C_{n,m}\int_{0}^{+\infty}x^{n+1}e^{-x^{2}}\left( L_{m}^{(n-1) }(x^2)\right) ^{2}\int_{t}^{+\infty }\lambda ^{-\frac{n}{2}}J_{n}\left(x\lambda ^{\frac{1}{2}}\right) d\lambda dx.  \label{4.29}
\end{equation}
Observe that the substitution $\lambda =ts$ transforms the
 last integral in \eqref{4.29} to
\begin{equation}
\int_{t}^{+\infty }\lambda ^{-\frac{n}{2}}J_{n}\left( x\lambda ^{%
\frac{1}{2}}\right) d\lambda =t^{1-\frac{n}{2}}\int_{1}^{+\infty }s^{-%
\frac{n}{2}}J_{n}\left( \left( xt^{\frac{1}{2}}\right) \sqrt{s}\right) ds,
\label{4.30}
\end{equation}
and by making use of formula (\cite[p. 691]{13})
\begin{equation}
\int_{1}^{+\infty }\rho ^{-\frac{n}{2}}(\rho -1)^{\mu-1}J_{\nu }(a\sqrt{\rho }) d\rho =\Gamma (\mu)
2^{\mu }a^{-\mu }J_{\nu -1}(a) ,  \label{4.31}
\end{equation}
$a>0$ and $0<\Re(\mu) <\frac{1}{2}\Re(\nu) +\frac{3}{4}$ for $a=xt^{\frac{1}{2}}$, $\mu =1$ and $\nu =n$, we obtain
\begin{equation}
\int_{t}^{+\infty }\lambda ^{-\frac{n}{2}}J_{n}\left( x\lambda ^{\frac{1}{2}}\right) d\lambda =2x^{-1}(\sqrt{t})^{1-n}J_{n-1}(x\sqrt{t}) .  \label{4.32}
\end{equation}
Consequently, the function $f_{m}(t) $ expressed by the formula
\eqref{4.29}, becomes:
\begin{equation}
f_{m}(t) =C_{n,m}t^{-\frac{n}{2}+\frac{1}{2}}\int_{0}^{+\infty }x^ne^{-x^{2}}J_{n-1}(x\sqrt{t}) \left(L_{m}^{n-1}(x^2) \right) ^{2}dx  \label{4.33}
\end{equation}
To calculate the integral \eqref{4.33}, we will discuss two cases: $n=1$ and $%
n\geq 2$.

For $n=1$, we make use the formula (\cite[p. 812]{13})
\begin{align}
\int_{0}^{+\infty }& x^{\nu +1}e^{-\alpha x^{2}}L_{p}^{(\nu-\sigma) }\left( \alpha x^{2}\right) L_{q}^{(\sigma)
}(\alpha x^{2}) J_{\nu }(xy) dx
\nonumber \\
&=\frac{\left( -1\right) ^{p+q}}{2\alpha }\left( \frac{y}{2\alpha }\right)
^{\nu }\exp \left( -\frac{y^{2}}{4\alpha }\right) L_{p}^{(\sigma-p+q) }\left( \frac{y^{2}}{4\alpha }\right) 
L_{q}^{(\nu -\sigma+p-q) }\left( \frac{y^{2}}{4\alpha }\right)   \label{4.34}
\end{align}
for $y=\sqrt{t},\alpha =1,\nu =0$ and $\sigma =0$. Then, we obtain that
\begin{equation}
f_{m}(t) =C_{1,m}\int_{0}^{+\infty}x^ne^{-x^{2}}J_{0}(x\sqrt{t}) L_{m}^{(0) }(x^{2}) L_{m}^{(0) }(x^2) dx=e^{-\frac{t}{4}%
}\left( L_{m}^{(0) }\left( \frac{t}{4}\right) \right) ^{2}.
\label{4.35}
\end{equation}
So, we get
\begin{equation}
B_{m}=\exp \left( \frac{1}{4}\Delta _{\C}\right) \left( L_{m}^{(0) }\left( -\frac{1}{4}\Delta _{\C}\right) \right) ^{2}.
\label{4.36}
\end{equation}
For $n\geq 2$, we first make use of the identity (\cite[p. 249]{3})
\begin{equation}
L_{p}^{(\alpha) }(y) =\sum\limits_{k=0}^{p}\frac{(\alpha -\beta) _{k}}{k!}L_{k-p}^{(\beta) }(y)   \label{4.37}
\end{equation}
for $p=m,\alpha =n-1,\beta =0$ and $y=x^{2}.$ Then, the equation \eqref{4.33}
becomes
\begin{equation}
f_{m}(t) =C_{n,m}t^{\frac{1-n}{2}}\sum\limits_{k=0}^{m}\frac{%
(n-1) _{k}}{k!}\int_{0}^{+\infty
}x^ne^{-x^{2}}L_{m}^{(n-1) }(x^2)
L_{m-k}^{(0) }(x^2) J_{n-1}(x\sqrt{t})
dx.  \label{4.38}
\end{equation}
To calculate the integral in the equation \eqref{4.38}, which is denoted
\begin{equation}
\mathcal{I}_{k}:=\int_{0}^{+\infty }x^ne^{-x^{2}}L_{m}^{(n-1) }(x^2) L_{m-k}^{(0) }(
x^{2}) J_{n-1}(x\sqrt{t}) dx,  \label{4.39}
\end{equation}
we return back to formula \eqref{4.34} and use it for $\nu =n-1,\sigma =0,p=m,q=m-k
$ and $y=\sqrt{t}$. We obtain that
\begin{equation}
\mathcal{I}_{k}=\frac{1}{2^n}t^{\frac{n-1}{2}}\exp \left( \frac{-t}{4}\right) 
L_{m}^{(-k) }\left( \frac{t}{4}\right) L_{m-k}^{(n+k-1) }\left( \frac{t}{4}\right) .  \label{4.40}
\end{equation}
Next, making use of the identity (\cite[p. 240]{3}),
\begin{equation}
L_{p}^{(-k) }(x) =(-x) ^{k}\frac{(p-k) !}{p!}L_{p-k}^{(k) }(x)  \label{4.41}
\end{equation}
for $p=m$ and $x=\frac{t}{4},$ the integral $\mathcal{I}_{k}$ becomes
\begin{equation}
\mathcal{I}_{k}=\frac{1}{2^n}\frac{(m-k)!}{m!}\left( -\frac{t}{4}\right) ^{k}t^{\frac{n-1}{2}}\exp \left( \frac{-t}{4}\right)
L_{m-k}^{(k) }\left( \frac{t}{4}\right) L_{m-k}^{(n+k-1) }\left( \frac{t}{4}\right)  \label{4.42}
\end{equation}
Then, we get that
\begin{align}
f_{m}(t) = &\frac{1}{(n) _{m}}\exp \left( \frac{-t}{4}\right) \nonumber
\\
& \times \sum\limits_{k=0}^{m}\frac{(n-1) _{k}(m-k) !}{%
k!}\left( -\frac{t}{4}\right) ^{k}L_{m-k}^{(k) }\left( \frac{t}{4}\right) L_{m-k}^{(n-1+k) }\left( \frac{t}{4}\right) .
\label{4.43}
\end{align}
Finally, we can write that
\begin{align}
B_{m}& =\frac{1}{(n) _{m}}\exp \left( \frac{1}{4}\Delta _{\C^n}\right)
\nonumber\\
&\times \sum\limits_{k=0}^{m}\frac{(n-1) _{k}( m-k)!}{%
k!}\left( \frac{\Delta _{\C^n}}{4}\right) ^{k}L_{m-k}^{(k) }\left( \frac{-\Delta _{\C^n}}{4}\right) L_{m-k}^{(n-1+k) }\left( \frac{-\Delta _{\C^n}}{4}\right)  \label{4.44}
\end{align}
We should note that the expression \eqref{4.44} enable us to rederive \eqref{4.36} when
replacing $n=1$ with the convention $(0)_{0}=1$. This help us
to summarize the discussion in one form as in the statement of the theorem.

%
%
%
%
%
%
%
%
%


\begin{thebibliography}{99}

\bibitem{1} F.A. Berezin, general concept of Quantization, Commn. math.
phys. 40, 153, Springer-Verlag 1975.

\bibitem{2} N. Askour, A. Intissar and Z. Mouayn, Espace de Bargmann
G\'{e}n\'{e}ralis\'{e}s et formules explicites pour leurs noyaux
reproduisants, C.R. Acad. Sci.paris,. t325, S\'{e}rie 1, P. 707-712, 1997.

\bibitem{3} W. Magnus, F. Oberhettinger and R. Soni, ''Formulas and Theorems for
Special functions of Mathematical Physics'', Springer-Verlag, Berlin
Heidelberg, New York (1966).

\bibitem{4} J. Peetre, The Berezin Transform and HA-Plitz Operator, J. Operator
Theory. 24, 165-186. 1990.

\bibitem{5} W. Rudin, Function theory in the unit Ball of $\C^n$,
Springer-Verlag, 1980.

\bibitem{6} N. Aronszain, Theory of Reproducing Kernels, Transaction of the American
Mathematical Society, Vol. 68, N$_{0}$ 3(May, 1950), pp.337-404.

\bibitem{7} N. Askour and Z. Mouayn, Spectral decomposition and Resolvent kernel for
the magnetic Laplacian in $\C^n$, Journal of Mathematical Physics.
41, Number 10, 6937-6943. 0ctober 2000.

\bibitem{8}  Z. Mouayn, Decomposition of magnetic Berezin transforms on the Euclidean
complex space $\C^n$, Integral Tranforms and special Functions, Vol.
19, N$_{0}$. 12, December 2008, 903-912.

\bibitem{9}  M.S. Birman and. Z. Solomjak, ''Spectral theory of Self-Adjoint Operator
in Hilbert Space'', D.Reidel publishing company, Dortrecht,Boston,
Lancaster, Tokyo (1987).

\bibitem{10}  V.S. Vladimirov, ''Equation of Mathematical Physics'', Mir Publishers
Moscow (1984).

\bibitem{11}  K. Yosida, ''Functional Analysis'', Springer-Verlag, Berlin Heidelberg,
New York (1968).

\bibitem{12}  V. Ditkine et A. Proudnikov, ''Transformations integrales et calcul
operationnel'', Edition Mir. Moscou (1978).

\bibitem{13}  I.S. Gradshteyn and I. M. Ryzhik, ''Table of Integrals, Series and
product'', Amsterdam, Elseuvier, seventh Edition 2007.

\end{thebibliography}
\end{document}